\documentclass{article}
\usepackage{amssymb}

\usepackage{amsmath}

\input{tcilatex}

\begin{document}

\title{Quantum Bi-Hamiltonian systems, alternative Hermitian structures and
Bi-Unitary transformations}
\author{\textbf{G. Marmo} \\
%EndAName
\textit{Dipartimento di Scienze Fisiche, Universita' di Napoli Federico II,
Complesso Universitario di Monte S Angelo, Napoli, Italy;} I.N.F.N., \textit{%
Sezione di Napoli, Italy}\\
giuseppe.marmo@na.infn.it \and \textbf{G. Scolarici} \\
%EndAName
\textit{Dipartimento di Fisica, Universita' di Lecce, 73100, Lecce;}
I.N.F.N., \textit{Sezione di Lecce, 73100 Lecce, Italy}\\
giuseppe.scolarici@le.infn.it \and \textbf{A. Simoni} \\
%EndAName
\textit{Dipartimento di Scienze Fisiche, Universita' di Napoli Federico II,
Complesso Universitario di Monte S Angelo, Napoli, Italy;} I.N.F.N., \textit{%
Sezione di Napoli, Italy}\\
alberto.simoni@na.infn.it \and \textbf{F. Ventriglia} \\
%EndAName
\textit{Dipartimento di Scienze Fisiche, Universita' di Napoli Federico II,
Complesso Universitario di Monte S Angelo, Napoli, Italy;} I.N.F.M \textit{%
and }I.N.F.N., \textit{Sezione di Napoli, Italy}\\
franco.ventriglia@na.infn.it}
\maketitle

\begin{abstract}
We discuss the dynamical quantum systems which turn out to be bi-unitary
with respect to the same alternative Hermitian structures in a
infinite-dimensional complex Hilbert space. We give a necessary and
sufficient condition so that the Hermitian structures are in generic
position. Finally the transformations of the bi-unitary group are explicitly
obtained.

\textbf{Keywords:} functional analysis, quantum mechanics, symplectic
geometry, spectral theory

\textbf{MSC 2000 classification: }58D19 Group actions and symmetry
properties; 37K05 Hamiltonian structures, symmetries, variational
principles, conservation laws.
\end{abstract}

\section{\protect\bigskip Introduction}

In the study of bi-Hamiltonian systems (see ref. \cite{ma} for quantum
systems) one starts with a given dynamics and looks for alternative
Hamiltonian descriptions. In this paper, we deal with the ''inverse
problem'' recently considered by some of us in different combinations \cite%
{msv}. We start with two Hermitian structures on complex Hilbert\ spaces
arising from two admissible triples in the corresponding real Hilbert space
and look for all dynamical quantum systems which turn out to be bi-unitary
with respect to them. However, we limit ourselwes to consider here the case
of admissible triples sharing the same complex structure. This study is a
preliminary step to generalize some results in Ref. \cite{mor}, to the
infinite-dimensional case.

This paper is organized as follows. In sec. 2 we show how to construct an
admissible Hermitian structure (admissible triple) on a real Hilbert space
starting from a metric tensor and a complex structure \cite{mor}. Moreover
we show how to recover a Hermitian scalar product after the complexification
of the Hilbert space. In sec. 3, we briefly discuss the bi-unitary group in
the case of finite dimensional spaces and relate the genericity condition of
the Hermitian structures with the cyclicity of the corresponding connecting
Hermitian operator. In sec. 4, we discuss bi-unitary groups in the
infinite-dimensional case, showing how the direct integral decomposition of
the Hilbert space with respect to a normed ring is a suitable theoretical
tool to deal with such a problem \cite{nai}. In particular we show how this
decomposition with respect to the bicommutant of the operator connecting two
Hermitian structures can be constructed and used for a full comparison of
the two Hermitian structures. In sec. 5, we prove that the component spaces
in the decomposition are unidimensional if and only if the Hermitian
structures are in generic position, what allows to conclude that all the
generic bi-unitary quantum systems commute. This construction resembles that
of the algebra one obtains for completely integrable systems by means of the
recursion operator \cite{def}. Moreover, the bi-unitary group is explicitly
exhibited both in the generic and non generic case. Finally, in the last
section, some concluding remarks are posed.

\section{Admissible Hermitian structures on the Hilbert space}

Let us show that given a metric tensor $g$ and a complex structure $J$ in a
infinite-dimensional real Hilbert space $\mathcal{H}^{\mathcal{R}}$, one can
construct an admissible Hermitian structure $h$ on $\mathcal{H}^{\mathcal{R}%
} $ which becomes an Hermitian scalar product in a corresponding complex
Hilbert space $\mathcal{H}$.

Given a couple $g$ and $J$ on $\mathcal{H}^{\mathcal{R}}$, we can always
construct out of them an admissible couple $g_{s}$, $J$ where $g_{s}$ is
defined in the following way

\begin{equation}
g_{s}=:\frac{1}{2}\{g(J.,J.)+g(.,.)\},
\end{equation}
which will be positive and nondegenerate if $g$ is positive and
nondegenerate. The metric $g_{s}$ and $J$ are admissible in the sense that $%
J $ is an anti-Hermitian operator with respect to $g_{s}$. Hereafter we drop 
$s $ in $g_{s}$. Such an admissible couple may be completed defining a
symplectic structure $\omega $ on $\mathcal{H}^{\mathcal{R}}$ as 
\begin{equation*}
\omega =g\circ J.
\end{equation*}
Then the triple $(g,J,\omega )$ is \textit{admissible.}

It is possible to obtain an admissible triple even starting from $g$ and $%
\omega $. In fact by Riesz's theorem a nonsingular linear operator $B$
exists such that

\begin{equation}
\omega (x,y)=g(Bx,y)
\end{equation}
and the antisymmetry of $\omega $ together with symmetry of $g$ imply that

\begin{equation}
g(Bx,y)=-g(x,By)
\end{equation}
so $B$ is skew-Hermitian and $-B^{2}>0$. Then a (symmetric) non-negative
square root of $B$, $R,$will be injective and densely defined. So $R^{-1}$
is well defined \cite{ab} and putting $J=:BR^{-1}$ and

\begin{equation}
g_{\omega }(.,.)=:g(R.,.)
\end{equation}
we recover

\begin{equation}
\omega (x,y)=g(Bx,y)=g_{\omega }(Jx,y)
\end{equation}
and in conclusion the triple $(g_{\omega },J,\omega )$ is admissible. In
other words $J$ is orthogonal and symplectic and also infinitesimally
orthogonal and infinitesimally symplectic, i.e. it is unitary and
skew-Hermitian: $J^{\dagger }=-J,$ $J^{2}=-\mathbf{1}$.

Now, from a metric tensor $g$ and an admissible symplectic form $\omega $,
an Hermitian structure on $\mathcal{H}^{\mathcal{R}}$ can be obtained. This
is a map $h:\mathcal{H}^{\mathcal{R}}\times \mathcal{H}^{\mathcal{R}%
}\rightarrow \mathbf{R}^{2}$ defined as follows

\begin{equation}
h:(x,y)\longmapsto (g(x,y),\omega (x,y)).
\end{equation}
Equivalently (and having in mind a quantum system) one can complexify $%
\mathcal{H}^{\mathcal{R}}$ by defining, for any complex number $z=\alpha
+i\beta $ and any vector $x$:

\begin{equation}
z.x=(\alpha +i\beta )x=:\alpha x+J\beta x.
\end{equation}

Then $h$ becomes an Hermitian scalar product, linear in the second factor,
on this complex Hilbert space $\mathcal{H}$, and we get

\begin{equation}
h(x,y)=g(x,y)+ig(Jx,y).
\end{equation}

\section{Bi-unitary descriptions in $\mathcal{H}$}

In quantum mechanics the Hilbert space $\mathcal{H}$ is given as a \emph{%
complex} vector space. Now we consider on $\mathcal{H}$ two different
Hermitian structures, coming from two admissible triples $%
(g_{1},J_{1},\omega _{1})$ and $(g_{2},J_{2},\omega _{2})$ on $\mathcal{H}^{%
\mathcal{R}}$ with the same complex structure: $J_{1}=J_{2}=J$. The
assumption $J_{1}=J_{2}$\ implies that the corresponding Hermitian
structures on $\mathcal{H}$ are compatible, in the sense that the group of
bi-unitary transformations is non void. \cite{mor}

Denoting with $h_{1}(.,.)$ and $h_{2}(.,.)$ the Hermitian structures given
on $\mathcal{H}$\ (both linear, for instance, in the second factor), we
search for the group that leaves both $h_{1}$ and $h_{2}$ invariant, that is
the bi-unitary group.

Again by Riesz's theorem a bounded, positive operator $G$ may be defined,
which is self-adjoint both with respect to $h_{1}$ and $h_{2}$, as

\begin{equation}
h_{2}(x,y)=h_{1}(Gx,y)\text{ \ \ \ }\forall x,y\in \mathcal{H}.
\end{equation}
Moreover, any bi-unitary operator $U$ must commute with $G$. Indeed

\begin{equation*}
h_{2}(x,U^{\dagger
}GUy)=h_{2}(Ux,GUy)=h_{1}(Ux,Uy)=h_{1}(x,y)=h_{2}(Gx,y)=h_{2}(x,Gy)
\end{equation*}
and from this

\begin{equation}
U^{\dagger }GU=G,\text{ \ \ }[G,U]=0.
\end{equation}
Therefore the group of bi-unitary operators belongs to the commutant $%
G^{\prime }$ of the operator $G$.

Let us discuss the bi-unitary group when $\mathcal{H}$ is
finite-dimensional. In this case $G$ is diagonalizable and the two Hermitian
structures result proportional in each eigenspace of $G$ \emph{via} the
eigenvalue. Then the group of bi-unitary transformations is

\begin{equation*}
U(n_{1})\times U(n_{2})\times ...\times U(n_{k})\text{ \ \ \ }%
n_{1}+n_{2}+...+n_{k}=n=\dim \mathcal{H},
\end{equation*}
where $n_{l}$ denotes the degeneracy of the l-th eigenvalue of $G$.

In finite-dimensional complex Hilbert spaces the following definition can be
stated \cite{mor}:

\textbf{Definition 1.} \textit{Two Hermitian forms are in generic position
iff the eigenvalues of }$G$\textit{\ are nondegenerate.}

Then, if \ $h_{1}$ and $h_{2}$ are in generic position the group of
bi-unitary transformations becomes

\begin{eqnarray*}
&&\underbrace{U(1)\times U(1)\times ...\times U(1)}. \\
&&\text{ \ \ \ \ \ \ \ \ \ \ \ }n\text{ factors}
\end{eqnarray*}

In other words this means that $G$ generates a complete set of observables.

Now we show that:

\textit{Two Hermitian forms are in generic position if and only if their
connecting operator }$G$\textit{\ is cyclic}.

In fact the non singular operator $G$ has a discrete spectrum and is
diagonalizable, so when $h_{1}$ and $h_{2}$ are in generic position $G$
admits $n$ dinstinct eigenvalues $\lambda _{k}$. Let now $\{e_{k}\}$ be the
eigenvector basis of $G$ and $\{\mu ^{k}\}$ an $n$-ple of nonzero complex
numbers. Then the vector 
\begin{equation}
x_{0}=\mu ^{k}e_{k}
\end{equation}
is a cyclic vector for $G$, that is the vectors $x_{0},$ $Gx_{0},...,$ $%
G^{n-1}x_{0}$ are $n$ linearly independent vectors. In fact one obtains 
\begin{equation}
G^{m}x_{0}=\mu ^{k}\lambda _{k}^{m}e_{k},
\end{equation}
and the coefficient determinant is given by 
\begin{equation}
(\prod\limits_{k}\mu ^{k})V(\lambda _{1},...,\lambda _{n}),
\end{equation}
where $V$ denotes the Vandermonde determinant which is different from zero
when all the eigenvalues $\lambda _{k}$ are distinct. The converse is also
true.

\section{Infinite-dimensional case}

Now we deal with the infinite-dimensional case, when the connecting operator 
$G$ may have a point part and a continuum part in its spectrum. As to the
point part, the bi-unitary group is $U(n_{1})\times U(n_{2})\times ...$
where now $n_{l}$ can also be $\infty $. When $G$ admits a continuum
spectrum, the charcterization of the bi-unitary group is more involved and
suitable mathematical tools are needed, such as spectral theory of operators
and theory of rings of operators in Hilbert spaces. In particular in the
infinite-dimensional case definition 1 of generic position of two Hermitian
forms is no more valid and has to be generalized in the following way \cite%
{mor}:

\textbf{Definition 2.} \textit{Two Hermitian forms are in generic position
iff }$G^{\prime \prime }=G^{\prime }$\textit{, that is when the bicommutant }%
$G^{\prime \prime }$\textit{\ coincides with the commutant }$G^{\prime }$%
\textit{\ of }$G$\textit{.}

First of all, observe that the commutant $G^{\prime }$\ and the bicommutant $%
G^{\prime \prime }$\ of the operator $G$ are weakly closed rings of bounded
operators in the ring of all bounded operators\ $\mathcal{B}(\mathcal{H})$\
on $\mathcal{H}$ and $G^{\prime \prime }\subset G^{\prime }$. Let $E_{0}$ be
the principal identity of a set $S\subset \mathcal{B}(\mathcal{H})$: by
definition $E_{0}$ is the projection operator on the orthogonal complement
of the set $KerS\cap KerS^{\dagger }$ and satisfies for all $A\in S$ the
following relation:

\begin{equation}
E_{0}A=AE_{0}=A.
\end{equation}

It can be proved \cite{nai} that the minimal weakly closed ring $R(S)$
containing $S$ is the strongest closure of the ring $R_{a^{\ast }}(S)$,
which is the minimal ring containing $S\cup S^{\dagger }$. $R(S)$ contains
only those elements $A\in S^{\prime \prime }$ which satisfy the above
condition (14).

Now, by the spectral theorem we express the selfadjoint operator $G$ in
terms of its spectral family $\{P(\lambda )\}$:

\begin{equation}
G=\int_{\Delta }\lambda dP(\lambda )
\end{equation}
where $\Delta =[a,b]$ is a closed interval containing the entire spectrum of 
$G$. Moreover, the positiveness of $G$ ensures that $KerG=0$. This implies
that $\mathbf{1}\in R(G)$ and hence $R(G)\equiv G^{\prime \prime }$.
Therefore $G^{\prime \prime }$ is commutative (in fact $R(G)$ is the
w-closure of the symmetric commutative subring $G$ in the ring of $\mathcal{B%
}(\mathcal{H})$).

We recall that each commutative weakly closed ring of operators $C$ in the
Hilbert space corresponds to a direct integral of Hilbert spaces.

The following theorems hold \cite{nai}:

\textbf{Theorem 1. }\textit{To each direct integral of Hilbert spaces}

\begin{equation*}
\mathcal{H}=\int_{\Delta }H_{\lambda }d\sigma (\lambda ),
\end{equation*}
\textit{with respect to a measure }$\sigma $\textit{\ there corresponds a
commutative weakly closed ring }$C=L_{\sigma }^{\infty }(\Delta )$\textit{\
where to each }$\varphi \in L_{\sigma }^{\infty }(\Delta )$\textit{\ there
corresponds the operator }$L_{\varphi }:(L_{\varphi }\xi )=\varphi (\lambda
)\xi _{\lambda }$ \textit{where} $\xi \in \mathcal{H},$ $\xi _{\lambda }\in
H_{\lambda }$\textit{\ and }$||L_{\varphi }||=||\varphi ||$\textit{.}

\emph{Vice versa}:

\textbf{Theorem 2. }\textit{To each commutative weakly closed ring }$C$%
\textit{\ of operators in a Hilbert space }$\mathcal{H}$\textit{\ there
corresponds a decomposition of }$\mathcal{H}$\textit{\ into a direct
integral, for which }$C$\textit{\ is the set of operators of the form }$%
L_{\varphi },$ $\varphi \in L^{\infty }$\textit{.}

From the previous theorems we get that \textit{the weakly closed commutative
ring }$R(G)$\textit{\ corresponds to a decomposition of the Hilbert space }$%
\mathcal{H}$ \textit{into the direct integral}

\begin{equation}
\mathcal{H}=\int_{\Delta }H_{\lambda }d\sigma (\lambda ),
\end{equation}
\textit{where} $\Delta =[a,b]$ \textit{is the entire spectrum of the
positive selfadjoint operator} $G$\textit{.}

The measure $\sigma (\lambda )$ is obtained by the spectral family $%
\{P(\lambda )\}$ and cyclic vectors in the usual way. Now every operator $A$
belonging to the commutant $G^{\prime }$ is representable in the form of a
direct integral of operators

\begin{equation}
A=\int_{\Delta }A(\lambda )\text{ }.\text{ }d\sigma (\lambda ),
\end{equation}
where $A(\lambda )$ is, for almost all $\lambda $, a bounded operator in $%
H_{\lambda }$.

Every operator $B$ of the bicommutant $G^{\prime \prime }=R(G)$ is then a
multiplication by a number $b(\lambda )$ on $H_{\lambda }$ for almost all $%
\lambda $.

Thus the bi-unitary transformations are in general a direct integral of
unitary operators $U(\lambda )$ acting on $H_{\lambda }$.

\section{Bi-unitary group transformations}

More insight can be gained by a more specific analysis of the direct
integral decomposition.

Let $G^{\prime }(\lambda )$ be the totality of all the operators $A(\lambda
) $ corresponding to $G^{\prime }$, for fixed $\lambda $. Then as $R(G)$ is
a maximal commutative ring in itself, we may conclude \cite{nai3} that the
family $G^{\prime }(\lambda )$ is irreducible for almost all $\lambda $, and
the direct integral of \ $\mathcal{H}\mathbb{\ }$with respect to $R(G)$ can
be written as follows

\begin{equation}
\mathcal{H}=\int_{\Delta }H_{\lambda }d\sigma (\lambda )=\sum_{k}\oplus 
\mathcal{H}_{k}=\sum_{k}\oplus \int_{\Delta _{k}}H_{\lambda }d\sigma
(\lambda ),
\end{equation}
where now the spectrum $\Delta $ decomposes into a sum of finite or
countable number of measurable sets $\Delta _{k}$, such that for $\lambda
\in \Delta _{k}$ the spaces $H_{\lambda }$ have constant dimension $k$ ($k$
can be $\infty $). Then any operator $A$ belonging to the commutant $%
G^{\prime }$ is representable as follows:

\begin{equation}
A=\sum_{k}\oplus \int_{\Delta _{k}}A(\lambda )\text{ }.\text{ }d\sigma
(\lambda ).
\end{equation}
As a consequence of Eq. (18), remembering that every operator $B$ of $%
G^{\prime \prime }=R(G)$ is a multiplication by a number $b(\lambda )$ on $%
H_{\lambda }$, we get the following result:

\textbf{Proposition 1. }\textit{Let two Hermitian structures} $h_{1}$ 
\textit{and} $h_{2}$ \textit{be given on the Hilbert space }$\mathcal{H}$%
\textit{. Then there exists a decomposition of }$\mathcal{H}$ \textit{into a
direct integral of Hilbert spaces }$H_{\lambda }$\textit{,} \textit{of
dimension }$k$\textit{,\ such that in each space }$H_{\lambda }$\textit{,\ }$%
h_{1}$ \textit{and} $h_{2}$\textit{\ are proportional: }$h_{2}=\lambda h_{1}$%
\textit{.}

Proposition 1 partially generalizes the Lemma in Ref. \cite{mor} to the case
of infinite-dimensional complex Hilbert spaces $\mathcal{H}$. This result
belongs to the constellation of propositions connected with the so called
''Quadratic Hamiltonian theorems''. \cite{suf}

As a consequence of Proposition 1, the elements $U$ of the bi-unitary group\
can be written as follows (see Eq. (19)):

\begin{equation}
U=\sum_{k}\oplus \int_{\Delta _{k}}U_{k}(\lambda )\text{ }.\text{ }d\sigma
(\lambda ),
\end{equation}
where $U_{k}:\Delta _{k}\rightarrow U(H_{\lambda })$, that is $U_{k}(\lambda
)$ is an element of the unitary group $U(H_{\lambda })$ on the spaces $%
H_{\lambda }$ of dimension $k$.

Moreover, when the Hermitian forms are in generic position the following
statement holds:

\textbf{Proposition 2.}\textit{\ The component spaces }$H_{\lambda }$\textit{%
\ of the decomposition of }$\mathcal{H}$\textit{\ into a direct integral\
with respect to }$R(G)$ \textit{are unidimensional if and only if the two
Hermitian forms }$h_{1}$ \textit{and} $h_{2}$\textit{\ are in generic
position.}

\textbf{Proof.} Let us suppose that two Hermitian forms are given in generic
position, then by definition $R(G)=G^{\prime \prime }=G^{\prime }$,
therefore $G^{\prime }$ must be commutative. Hence the totality of all
irreducibles operators $A(\lambda )$ corresponding to $G^{\prime }$ are
unidimensional, then $\dim H_{\lambda }=1$ for all $\lambda \in \Delta $.

In order to prove the converse, observe that if $R(G)=G^{\prime \prime }\neq
G^{\prime }$, there exists a non zero subset $\Delta _{0}$ of $\Delta $ such
that for $\lambda \in \Delta _{0}$ the totality of irreducible operators $%
A(\lambda )$ corresponding to $G^{\prime }$ is not commutative. Hence $\dim
H_{\lambda }\neq 1.$ $\blacksquare $

Proposition 2 fully generalizes the Proposition in Ref. \cite{mor} to the
case of infinite-dimensional complex Hilbert space, so that we can say that
all the generic different admissible quantum dynamical systems are pairwise
commuting.

Finally, Proposition 2 implies that the unitary operators $U_{k}(\lambda )$
in Eq. (20), reduce to a multiplication by a phase factor $e^{i\varphi
(\lambda )}$ on $H_{\lambda }$ for almost all $\lambda $, so that the
elements of the bi-unitary group read

\begin{equation}
U=\int_{\Delta }e^{i\varphi (\lambda )}\text{ }.\text{ }d\sigma (\lambda ).
\end{equation}

\section{Concluding remarks}

In this paper we have shown how to exend to the more realistic case of
infinite dimensions our results of a previous paper dealing mainly with
finite level quantum systems.

Our approach shows how to deal with ''pencils of compatible Poisson
Brackets'' \cite{ge} in the framework of quantum systems.

Our formulation is already in a form suitable to deal with quantum
electrodynamics.

We hope to be able to extend these results to the evolutionary equations for
classical and quantum field theories.


\begin{thebibliography}{9}
\bibitem{ma} F. Magri, \textit{A simple model of integrable Hamiltonian
equation, }J. Math. Phys. \textbf{19}, 1156-1162 (1978); F. Magri, \textit{A
geometrical approach to the nonlinear solvable equations, }Lect. Notes in
Phys. \textbf{120}, 233-263 (1980); B. Fuchssteiner, \textit{The Lie algebra
structure and degenerate Hamiltonian and bi-Hamiltonian systems, }Progr.
Theor. Phys.\textit{\ }\textbf{68},1082-1104 (1982).

\bibitem{msv} F. Ventriglia, \textit{Alternative Hamiltonian descriptions
for quantum systems and non-Hermitian operators with real spectrum, }Mod.
Phys. Lett.\textit{\ }\textbf{A 17}, 1589-1599 (2002); G. Marmo, A. Simoni
and F. Ventriglia, \textit{Quantum systems: real spectra and nonhermitian
(Hamiltonian) operators, }Rep. Math. Phys., \textbf{51}, 275-285 (2003); G.
Marmo, A. Simoni and F. Ventriglia, \textit{Quantum systems and alternative
unitary descriptions,} Int. J. Mod. Phys. A, to appear (2004).

\bibitem{mor} G. Marmo, G. Morandi, A. Simoni and F. Ventriglia, \textit{%
Alternative structures and bi-Hamiltonian systems, }J. Phys. A: Math. Gen., 
\textbf{35}, 8393-8406 (2002).

\bibitem{nai} M. A. Naimark:\textit{\ }Normed Rings,\textit{\ }%
Wolters-Noordhoff \ Publishing Groningen, 1970.

\bibitem{def} S. De Filippo, G. Marmo, M. Salerno and G. Vilasi, \textit{A
new characterization of completely integrable systems, }Nuovo Cimento\textit{%
\ }\textbf{B 83} 97-112 (1984).

\bibitem{ab} R. Abraham and J. E. Marsden: Foundation of Mechanics,
Benjamin-Cummings, New York, 1978.

\bibitem{nai3} See Ref. \cite{nai} p. 503.

\bibitem{suf} G. Marmo, E. J. Saletan, R. Schmid and A. Simoni, \textit{%
Bi-Hamiltonian Dynamical Systems and the Quadratic-Hamiltonian Theorem, }%
Nuovo Cimento\textit{\ }\textbf{B 100} 297-317 (1987).

\bibitem{ge} I. M. Gel'fand and I Zakharovich: On Local Geometry of
Bihamiltonian Structure\textit{\ }(Gel'fand Mathematical Seminar vol 1) ed
L. Corwin, I. M. Gel'fand and J. Lepowski, Birkauser, Basel 1993; A. Ibort,
F. Magri and G. Marmo, \textit{Bihamiltonian structures and Stackel
separability, }J. Geom. Phys.\textit{\ }\textbf{33} 210-228 (2000).
\end{thebibliography}
\end{document}